\documentclass[10pt,english,english,english,english]{article}
\usepackage[T1]{fontenc}
\usepackage[latin1]{inputenc}
\usepackage{babel}
\usepackage{graphics}

\makeatletter

\providecommand{\LyX}{L\kern-.1667em\lower.25em\hbox{Y}\kern-.125emX\@}
\newcommand{\noun}[1]{\textsc{#1}}
\let\SF@@footnote\footnote
\def\footnote{\ifx\protect\@typeset@protect
    \expandafter\SF@@footnote
  \else
    \expandafter\SF@gobble@opt
  \fi
}
\expandafter\def\csname SF@gobble@opt \endcsname{\@ifnextchar[%]
  \SF@gobble@twobracket
  \@gobble
}
\edef\SF@gobble@opt{\noexpand\protect
  \expandafter\noexpand\csname SF@gobble@opt \endcsname}
\def\SF@gobble@twobracket[#1]#2{}

 \newcommand{\lyxaddress}[1]{
   \par {\raggedright #1 
   \vspace{1.4em}
   \noindent\par}
 }

\usepackage[T1]{fontenc}
\usepackage[latin1]{inputenc}
\usepackage{babel}
\usepackage{graphics}

\makeatletter

\usepackage[T1]{fontenc}
\usepackage[latin1]{inputenc}
\usepackage{babel}
\usepackage{graphics}
\usepackage{longtable}

\makeatletter

\usepackage[T1]{fontenc}
\usepackage[latin1]{inputenc}
\usepackage{babel}
\usepackage{graphics}
\usepackage{longtable}



\makeatother

\makeatother
\begin{document}

\title{\textbf{Consequences of the extra SM Families on the Higgs Boson
Production at Tevatron and LHC }}

\author{E. Arik\(^{1}\), O. \c{C}akir\(^{2}\), S.A. \c{C}etin\( ^{1} \), 
S. Sultansoy\( ^{3,4} \)}

\maketitle

\lyxaddress{\( ^{1} \)Department~of~ Physics,~ Faculty~ of~ Arts~ and~
Sciences,~Bogazici~University,~80815,~ Bebek, Istanbul,~ Turkey.}

\lyxaddress{\( ^{2} \)Department~ of~Physics,~ Faculty~ of~ Sciences,~
Ankara University,~ 06100, Tandogan,~ Ankara,~ Turkey.}

\lyxaddress{\( ^{3} \)Department~ of~ Physics,~ Faculty~ of~ Arts~ and~
Sciences,~ Gazi~ University,~ 06500,~Teknikokullar, Ankara,~Turkey.}

\lyxaddress{\( ^{4} \)Institute~ of~ Physics,~ Academy~ of~ Sciences,~
H.~ Cavid~ Avenue 33,~ 370143, Baku,~ Azerbaijan.}

\begin{abstract}
The latest electroweak precission data allow the existence of additional
chiral generations in the standard model. We study the influence of
extra generations on the production of SM Higgs boson at hadron colliders.
Due to the enhancement of the gluon fusion channel, the {}``golden
mode'' becomes more promising even at upgraded Tevatron. Furthermore,
the formation of the fourth family quarkonia with the subsequent \textbf{\( \eta \rightarrow Zh \)}
decay introduces additional tool for the investigation of the Higgs
boson properties. 
\end{abstract}
Two years ago, the existence of extra Standard Model (SM) generations
seemed to be \char`\"{}excluded\char`\"{} by precission electroweak
data \cite{erler}. However, with the latest electroweak data, the
situation is changed. In a recent paper \cite{novikov}, it is shown
that the quality of the fit for one new generation is as good as for
zero new generation; moreover, two and even three partially heavy
generations are allowed when neutral fermions are relatively light,
\( m_{N}\approx 50 \) GeV. Using LEP2 data for the process \( e^{+}e^{-}\rightarrow Z^{\star }\rightarrow N\overline{N}\gamma  \)
one can exclude three partially heavy generations which contain such
a light \( N \) at a level of \( 3\sigma  \) while one or even two
such generations may exist \cite{ilyin}. Then, according to \cite{he}
a single extra chiral family with a constrained spectrum is consistent
with precision data without requiring any other new physics source,
and the precision bound on the SM-like Higgs boson mass is significantly
relaxed in the presence of an extra relatively light chiral family,
namely, SM Higgs mass up to about \( 500 \) GeV is allowed.

Direct searches for new leptons (\(N,\textrm{ }L \)) and quarks
(\(u_{4},\textrm{ }d_{4} \)) lead to the following lower bounds
on their masses \cite{groom}: \( m_{L}>92.4 \) GeV, \( m_{N}>45 \)
GeV (Dirac), \( m_{N}>39.5 \) GeV (Majorana), \( m_{d_{4}}>199 \)
GeV (neutral current decays), \( m_{d_{4}}>128 \) GeV (charged current
decays).

It is known that the SM does not predict the number of generations.
In the Democratic Mass Matrix approach, the SM is extended to include
extra generations (see \cite{sultansoy} and references therein).

The leading production mechanism for a SM Higgs boson at hadron colliders
is the gluon-fusion process via heavy quark triangle loop

\begin{equation}
p\overline{p}\rightarrow ggX\rightarrow hX.
\end{equation}
 To the lowest order, cross section is given by

\begin{equation}
\sigma (p\bar{p}\rightarrow hX)=\sigma _{0}\tau _{h}\int _{\tau _{h}}^{1}{dx\over x}g(x,Q^{2})g(\tau _{h}/x,Q^{2})
\end{equation}
 where \( g(x,Q^{2}) \) denotes the gluon density of proton and \( \tau _{h}=m_{h}^{2}/s \).
The natural values are chosen for the factorization scale \( Q^{2} \)
(\( =m_{h}^{2} \)) of the parton densities and the renormalization
scale \( \mu  \)(\( =m_{h} \)) for the running strong coupling constant
\( \alpha _{s}(\mu ) \). The partonic cross section is given by \cite{spira95}

\begin{equation}
\sigma _{0}(gg\rightarrow h)={G_{F}\alpha _{s}^{2}(\mu ^{2})\over \sqrt{2}\, 288\pi }\sum _{Q}|g_{Q}A_{Q}|^{2}.
\end{equation}
 The expression for the amplitude \( A_{Q} \) can be found in \cite{spira95}.
The Yukawa coupling is given by \( g_{Q}=m_{Q}/v \), where \( v=246 \)
GeV is the vacuum expectation value of the Higgs field and \( m_{Q} \)
denotes the heavy quark mass. In spite of the large value of the fourth
SM family quark masses, perturbation theory is still applicable for
\( m_{Q}<870 \) GeV since \( \alpha _{Q}=g_{Q}^{2}/4\pi <1 \). This
restriction should be considered as conservative because the scale
of the corrections to Yukawa interactions of the Higgs boson with
heavy quarks is given by \( g_{Q}^{2}/(4\pi )^{2} \) and these corrections
can be neglected for heavy quark masses upto \textbf{\( m_{Q}\approx  \)}
3 TeV \cite{ginzburg99}.

\begin{figure}
\centering \resizebox*{8cm}{8cm}{\includegraphics{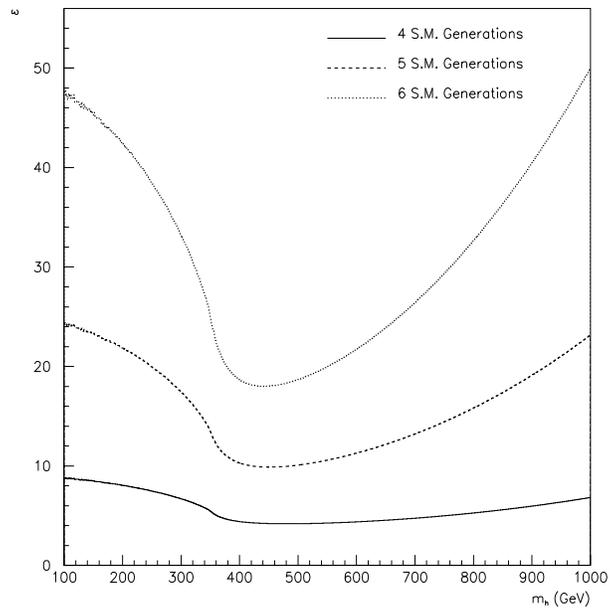}  }
\caption{Enhancement factor \protect\protect\protect\( \epsilon \protect \protect \protect \)
as a function of Higgs boson mass, where quarks from extra generations
are assumed to be infinitely heavy. \label{fig1}}
\end{figure}

Quarks from extra generations contribute to the loop mediated process
in Higgs boson production \( gg\rightarrow h \) at hadron colliders
resulting in an enhancement of \( \sigma _{0} \) by a factor of \textbf{\( \epsilon  \)}.
It is obvious that in the infinitely heavy quark mass limit \cite{dawson},
the expected enhancement factors are \( \epsilon = \)9, 25 and 49
in the case of one, two and three extra generations, respectively.
In Fig. \ref{fig1}, we plotted \( \epsilon  \) as a function of
\( m_{h} \) assuming quarks from extra generations to be infinitely
heavy.

\begin{figure}
\centering \resizebox*{6cm}{6cm}{\includegraphics{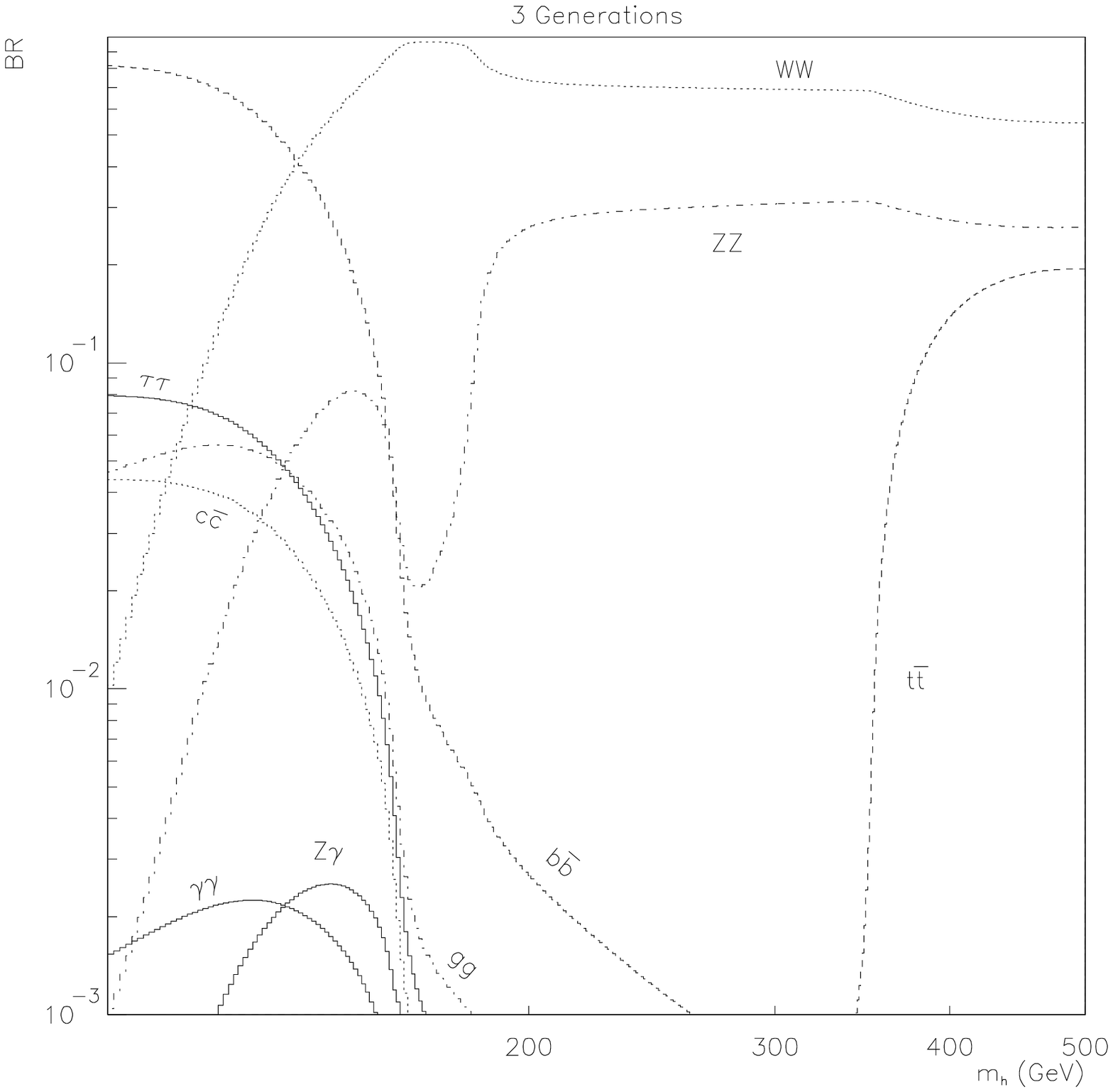}  } \resizebox*{6cm}{6cm}{\includegraphics{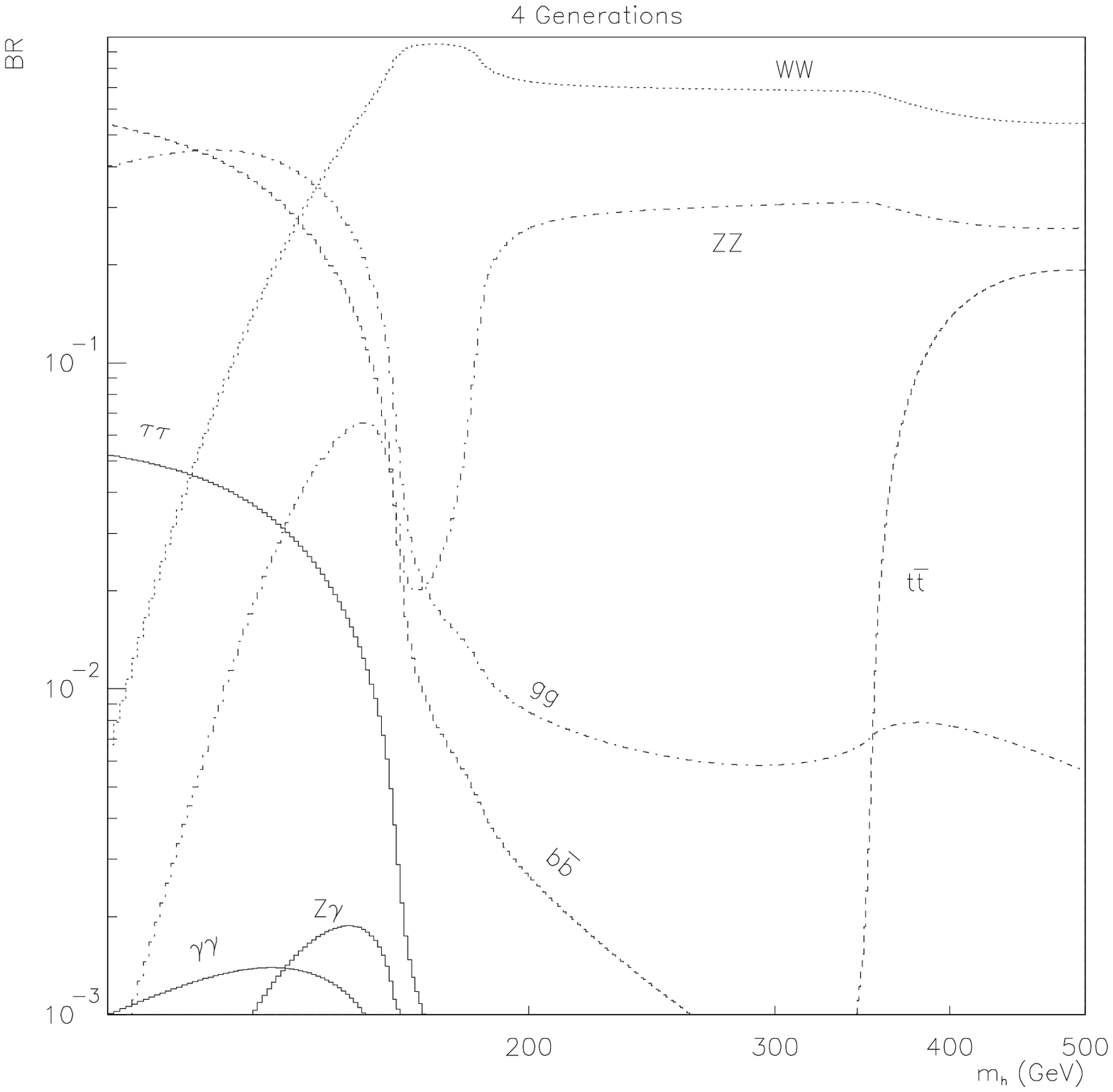} 
} \centering \resizebox*{6cm}{6cm}{\includegraphics{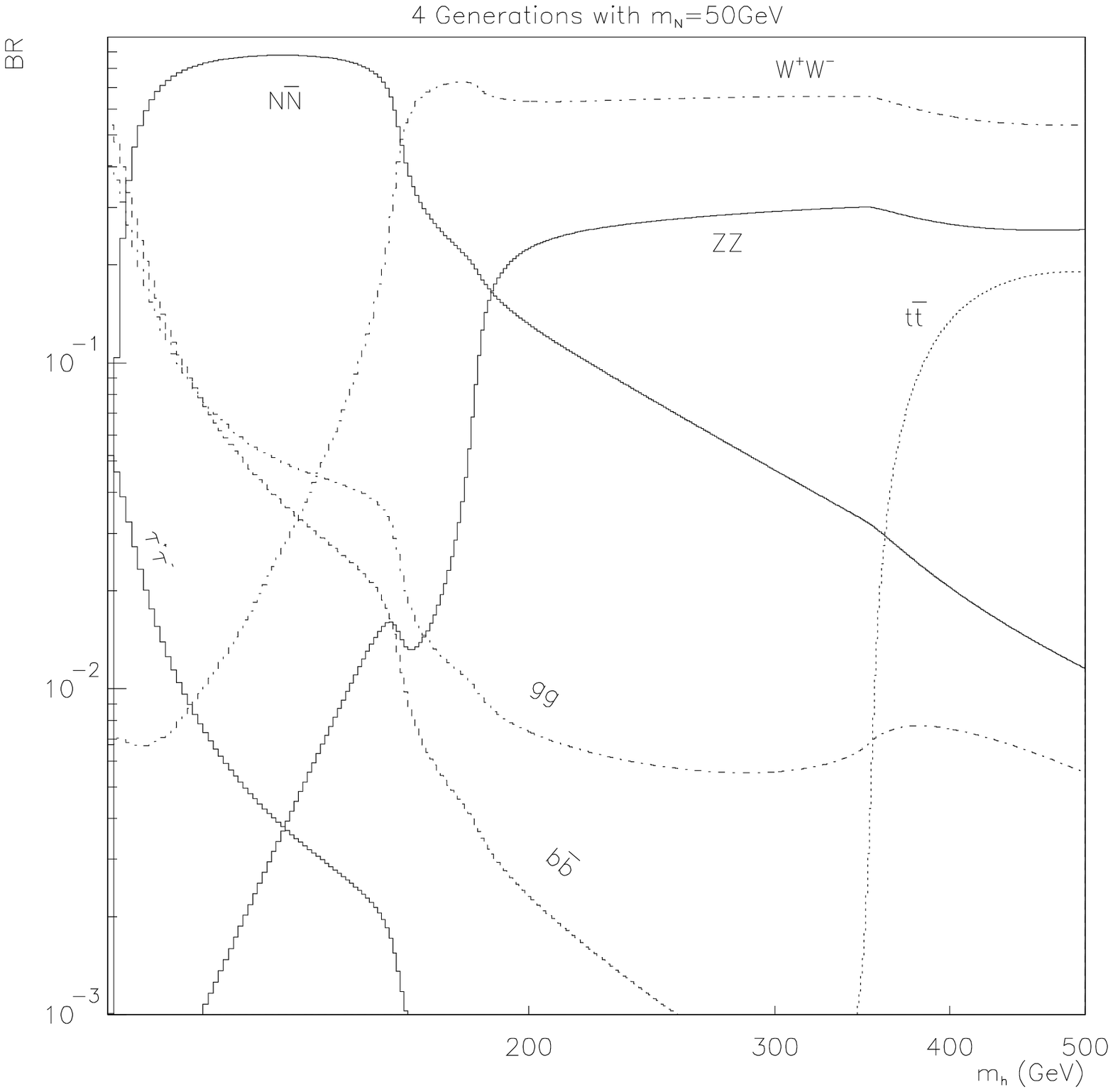}  } \resizebox*{6cm}{6cm}{\includegraphics{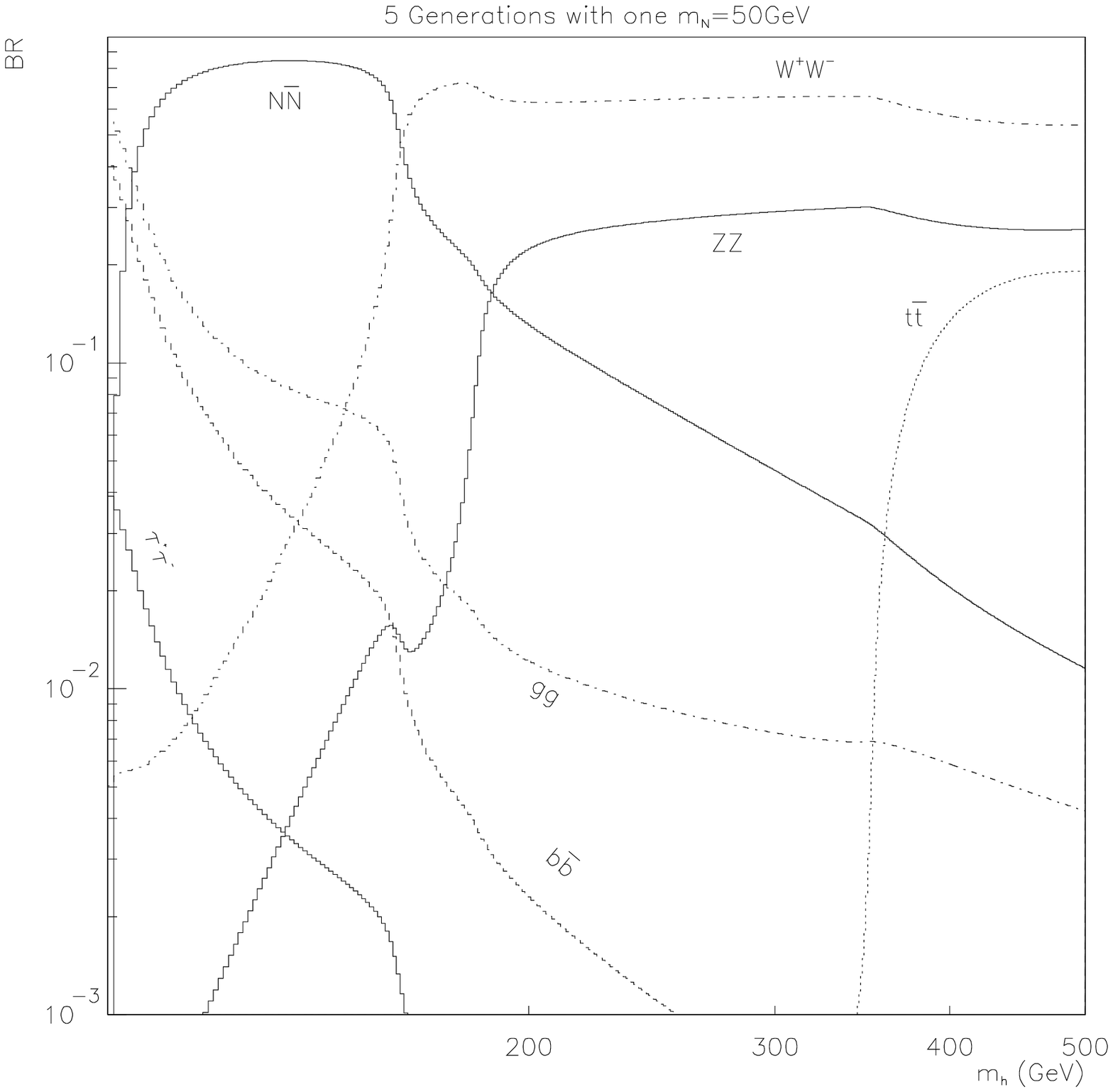} 
}
\caption{Branching ratios of the main decay modes of the SM Higgs boson. \label{fig2}}
\end{figure}

Extra generations will also effect \( h\rightarrow \gamma \gamma  \),
\( Z\gamma  \) decay widths. In addition, if \( m_{h}>2m_{N} \)
a new channel, namely \( h\rightarrow N\overline{N} \) appears. Branching
ratios of the main decay modes of the Higgs boson are plotted in Fig.
\ref{fig2} for zero, one and two extra SM generations. In our calculations,
we use \cite{spira95} with minor modification\textbf{s} due to extra
SM generations. According to \cite{lorenzo}, extra generations will
give a negative corrections to the \( hZZ \) vertex. Since the same
corrections take place also for the \( hWW \) vertex, this effect
can be neglected for \( m_{h}>150 \) GeV where \( h\rightarrow WW^{(\star )} \)
channel is dominant. For \( m_{h}<150 \) GeV these corrections can
be neglected for relatively light extra generations (\( m_{Q}<300 \)
GeV).

In this study, we concentrate on the golden mode \( h\rightarrow ZZ^{(\star )}\rightarrow 4l \),
where \( l=e,\mu  \). In Table \ref{table1} we present production
cross section for golden events in the cases of three, four, five
and six SM generations at Tevatron with \( \sqrt{s}=2 \) TeV. The
cross sections for LHC with \( \sqrt{s}=14 \) TeV are given in Table
2. In our calculations we use PYTHIA 6.2 \cite{PYTHIA} with the CTEQ5L
parton distribution \cite{CTEQ} and have normalized our signal cross
section to include NLO QCD corrections \cite{Graduenz}.

\begin{figure}
\centering \resizebox*{10cm}{10cm}{\includegraphics{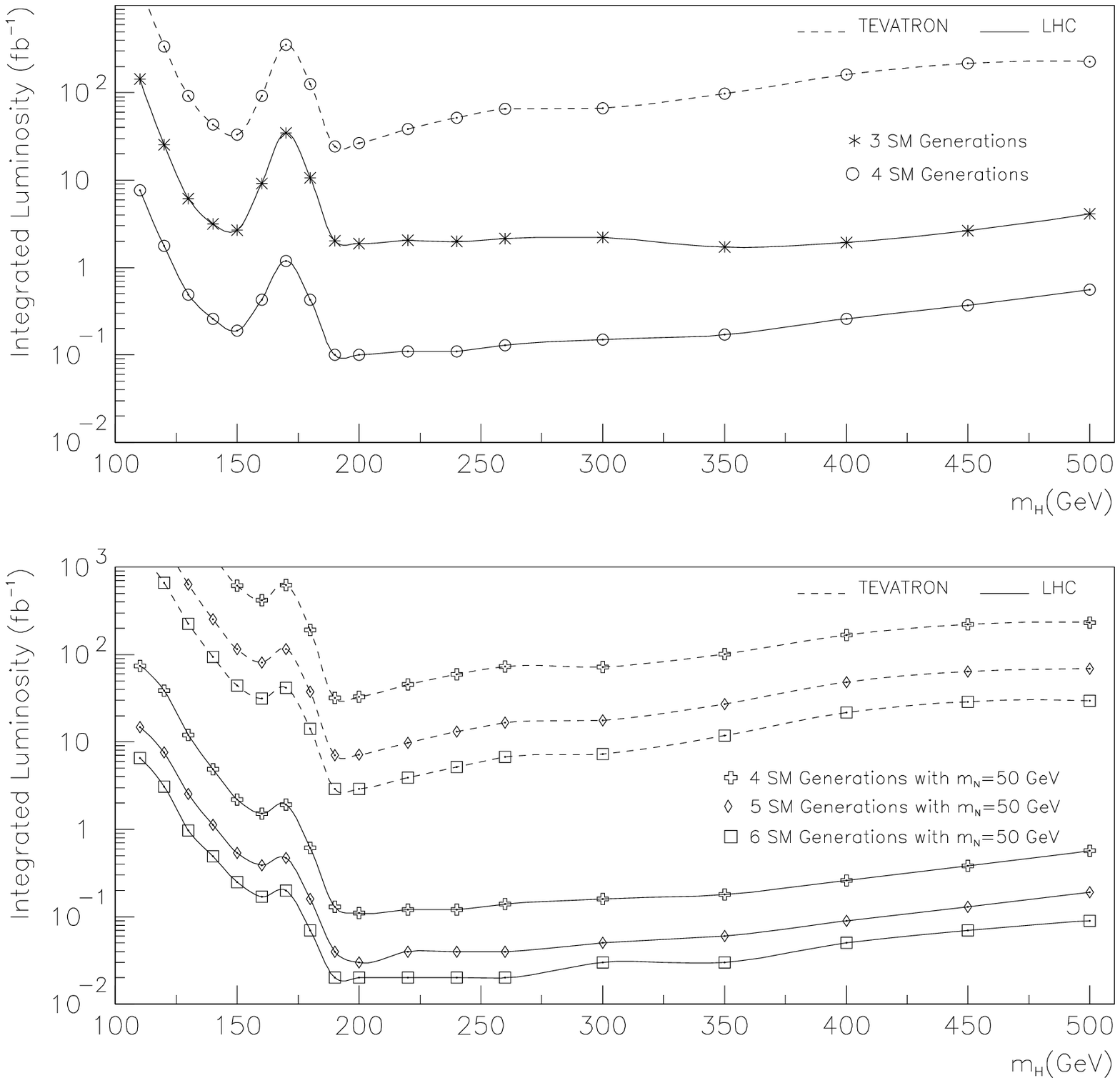}  }
\caption{The luminosity needed to achieve \protect\protect\protect\( 3\sigma \protect \protect \protect \)
significance for golden mode at Tevatron and LHC. \label{fig3}}
\end{figure}

\begin{figure}
\centering \resizebox*{10cm}{10cm}{\includegraphics{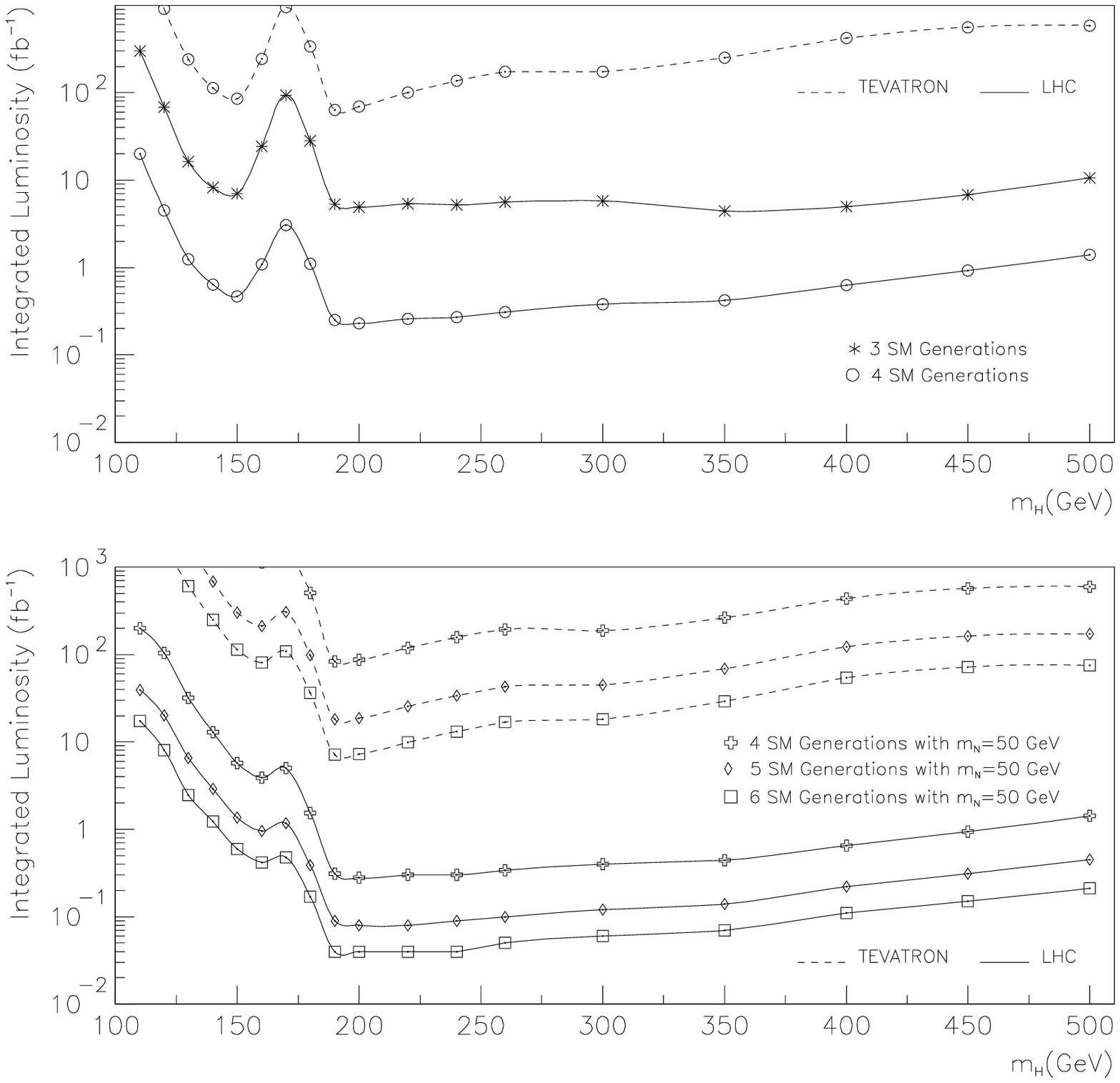}  }
\caption{The luminosity needed to achieve \protect\protect\protect\( 5\sigma \protect \protect \protect \)
significance for golden mode at Tevatron and LHC. \label{fig4}}
\end{figure}

The most important background is the pair production of Z bosons,
\( p\overline{p}(pp)\rightarrow ZZ^{(\star )}X \) \( (Z\rightarrow l^{+}l^{-}) \),
which has \( \sigma \approx 8 \) fb at Tevatron and \( \sigma \approx 45 \)
fb at LHC. Since the Higgs boson width grows rapidly with \( m_{h} \),
a variable mass window of width \( \sigma _{m} \) (given by the convolution
of the Higgs decay width \( \Gamma _{h} \) and of the experimental
resolution estimated to be 2\% of the Higgs boson mass)

\begin{equation}
\sigma _{m}=\sqrt{\left( \frac{\Gamma _{h}}{2.36}\right) ^{2}+\left( 0.02m_{h}\right) ^{2}}
\end{equation}
 is used in order to estimate the numbers of signal and background
events. In this case, the acceptance for signal events is \( 90\% \)
in a mass window of \( \pm 1.64\sigma _{m} \) around \( m_{h} \).
The integral luminosities needed to achieve a \( 3\sigma  \) (\( 5\sigma  \))
significance level of Higgs boson observation at Tevatron and LHC
are shown in Fig. \ref{fig3} (\ref{fig4}). In our calculations we
assume rather conservative value of 25\% for the overall acceptance
in four lepton final states, and the integrated luminosities to achieve
the desired statistical significances are obtained using Poisson statistics.

From Fig. \ref{fig3}, one can see that Higgs boson with mass around
150 and 200 GeV will be observed at upgraded Tevatron with \( L_{int}=30 \)
fb\( ^{-1} \) if the fourth SM family exist\textbf{s}. In this case,
Higgs boson with \( 110<m_{h}<500 \) GeV will be seen at LHC via
golden mode even at low luminosity of \( L_{int}=10 \) fb\( ^{-1} \).
In the case of partially heavy extra generations, \( m_{N}=50 \)
GeV, Higgs boson with mass \( 180<m_{h}<350 \) GeV (\( 170<m_{h}<450 \)
GeV) will be observed at a \( 3\sigma  \) level at upgraded Tevatron
if there are five (six) SM generations. In other words, Tevatron can
exclude two (three) extra generations at \( 3\sigma  \) level if
Higgs mass lies between 180 (170) GeV and 350 (450) GeV and golden
mode is not seen.

Fig. \ref{fig4} shows that two extra generations can be excluded
at \( 5\sigma  \) level by upgraded Tevatron (before LHC start) if
Higgs mass lies between 180 and 250 GeV. Three extra generations will
be excluded for \( 175<m_{h}<350 \) GeV. On the other hand, LHC with
moderate integrated luminosity of 10 fb\( ^{-1} \), will exclude
(or confirm) degenerate extra SM generations via golden mode practically
for the whole region of Higgs masses. In the case of partially heavy
extra generations and \( m_{h}<140 \) GeV, higher integrated luminosity
(up to 100 fb\( ^{-1} \)) is needed.

A non degenerate multiplet of heavy extra fermions will affect the
parameter \( \rho  \) \cite{erler}. For example, assuming degenerate
fourth generation quarks, one can easily obtain a condition for 
non-degenerate
leptons: 
\begin{equation}
1+x^{2}-\frac{4x^{2}}{1-x^{2}}\textrm{ln}\frac{1}{x}\leq \frac{3\times 10^{4}}{m_{N}^{2}}
\end{equation}
where \( x=m_{L}/m_{N} \). Therefore, fourth generation charged
lepton mass limits are \( m_{L}\leq 208 \) GeV and \( 350\leq m_{L}\leq 650 \)
GeV for \( m_{N}=50 \) GeV and \( m_{N}=500 \) GeV, respectively.

According to \cite{novikov} in the case of partially heavy extra
generations with \( m_{N}\simeq 50 \) GeV the preferable mass value
of the fourth generation charged lepton is around \( 130 \) GeV.
For the Higgs mass larger than \( 2m_{L} \) the decay channel \( h\rightarrow L^{+}L^{-} \)
will contribute to Higgs decay and branchings. Concerning the golden
mode this contribution will decrease the branching ratio approximately
\( 5\% \). Our results will not change for \( m_{h}<2m_{L} \) and
slightly change for \( m_{h}>2m_{L}. \)

\begin{figure}
\centering \resizebox*{8cm}{8cm}{\includegraphics{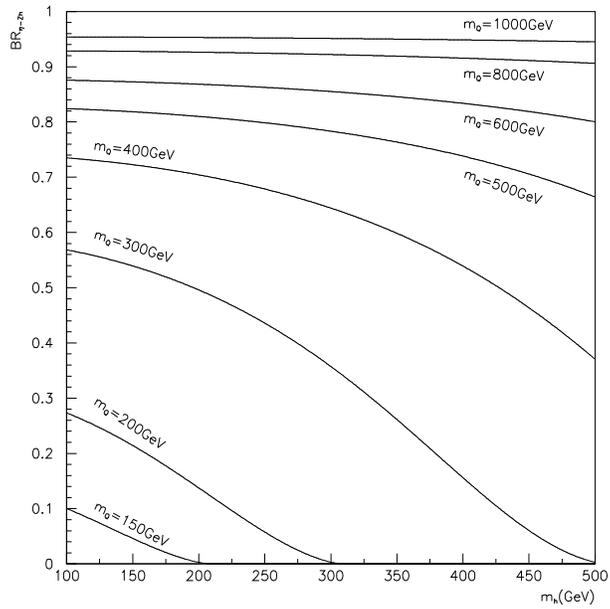}  }
\caption{Branching ratio for the process \protect\protect\protect\( \eta \rightarrow Zh\protect \protect \protect \).
\label{fig5}}
\end{figure}

\begin{figure}
\centering \resizebox*{8cm}{8cm}{\includegraphics{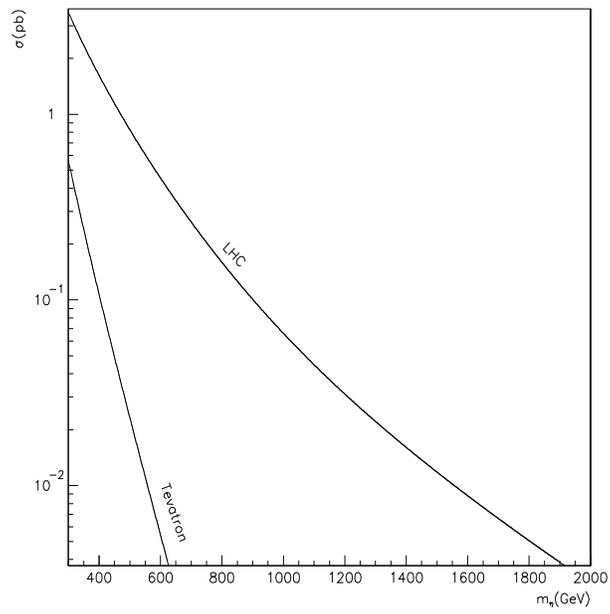}  }
\caption{Production cross section for \protect\protect\protect\( \eta \protect \protect \protect \)
quarkonia. \label{fig6}}
\end{figure}

An additional opportunity for investigation of Higgs boson properties
at hadron colliders appears if the quarks from extra generations form
a quarkonia. The condition for forming (\( Q\overline{Q} \)) quarkonia
states with new heavy quarks is \cite{bigi}:

\begin{equation}
m_{Q}\leq (125\textrm{ GeV})|V_{Qq}|^{-2/3}
\end{equation}
 where \( Q \) and \( \textrm{q} \) denote new heavy quarks and
known quarks, respectively; \( V_{Qq} \) is the corresponding Cabibbo-Kobayashi-Maskawa
(CKM) matrix element. In the case of four SM generations this condition
is satisfied by parametrization given in \cite{celikel,atag}. The
Higgs boson will be produced via the subprocess \( gg\rightarrow \eta (Q\overline{Q})\rightarrow Zh \).
In Fig. \ref{fig5}, we plot branching ratio \( BR(\eta \rightarrow Zh) \)
as a function of \( m_{h} \) for different values of the masses of
the fourth SM generation quarks. Here, we use corresponding formulas
from \cite{barger} in the framework of Coulomb potential model. Production
cross sections for \textbf{\( \eta  \)} quarkonia at Tevatron and
LHC are plotted in Fig. \ref{fig6}. At Tevatron, it is obvious that
the considered mechanism is a matter of interest only for relatively
light new quarks. For example, if \( m_{\eta }=400 \) GeV and \( m_{h}=150 \)
GeV, one expects 600 \( Zh \) events at \( L_{int}=30 \) fb\( ^{-1} \).
This channel seems to be much more promising at LHC: with the same
statements we expect more than \textbf{\( 3\times 10^{4} \)} \( Zh \)
events for \( L_{int}=100 \) fb\( ^{-1} \). More comprehensive results
on the subject including the effects of specific mass patterns on
quarkonia decays will be reported elsewhere \cite{arik}.

In conclusion, existence of the extra generations will significantly
affect Higgs boson production at hadron colliders. In this paper,
we examined the golden mode \( h\rightarrow ZZ^{(\star )}\rightarrow 4l \).
Of course, the same statement is valid for other decay channels, too.
For example, the modes \textbf{\( h\rightarrow WW^{(\star )}\rightarrow l\nu jj \)}
and \textbf{\( h\rightarrow WW^{(\star )}\rightarrow l\overline{\nu }\overline{l}\nu  \)}
are very promising at Tevatron for \( 135<m_{h}<180 \) GeV \cite{han,turcot}.
Another promising channel for Higgs boson production is the formation
of pseudoscalar \textbf{\( \eta  \)} quarkonia with subsequent decay
\( \eta \rightarrow Zh \). Finally, both upgraded Tevatron and LHC
will give opportunity to determine the actual number of SM generations.

\pagebreak

\noun{Acknowledgements}

We would like to thank A. Celikel, A.K. Ciftci and R. Ciftci for useful
discussions. We are also grateful to H.-J. He and J. Lorenzo Diaz-Cruz
for useful remarks. This work is partially supported by Turkish Planning
Organization (DPT) under the Grant No 2002K120250.

\newpage

\begin{table}

\caption{The production cross sections in fb for golden events at Tevatron.
The asterix {*} denotes that the calculations are performed assuming
one neutrino with \protect\protect\protect\protect\( m_{N}=50\protect \protect \protect \protect \)
GeV. \label{table1}}

\begin{longtable}{cccccc}
\hline 
\( m_{h} \)(GeV)&
SM-3&
SM-4&
SM-4{*}&
SM-5{*}&
SM-6{*}\\
\hline 
100&
0.003&
0.019&
0.019&
0.040&
0.051\\
110&
0.010&
0.055&
0.014&
0.036&
0.060\\
120&
0.030&
0.157&
0.023&
0.061&
0.108\\
130&
0.061&
0.332&
0.041&
0.108&
0.194\\
140&
0.085&
0.500&
0.064&
0.170&
0.309\\
150&
0.086&
0.577&
0.098&
0.262&
0.480\\
160&
0.036&
0.283&
0.114&
0.307&
0.571\\
170&
0.016&
0.128&
0.092&
0.249&
0.476\\
180&
0.033&
0.271&
0.213&
0.578&
1.103\\
190&
0.115&
0.932&
0.787&
2.129&
4.084\\
200&
0.116&
0.926&
0.805&
2.176&
4.173\\
220&
0.088&
0.686&
0.617&
1.659&
3.178\\
240&
0.068&
0.511&
0.469&
1.254&
2.396\\
260&
0.052&
0.383&
0.357&
0.948&
1.808\\
300&
0.047&
0.314&
0.300&
0.779&
1.474\\
350&
0.037&
0.201&
0.195&
0.481&
0.889\\
400&
0.027&
0.117&
0.114&
0.268&
0.484\\
450&
0.019&
0.081&
0.080&
0.189&
0.341\\
500&
0.016&
0.066&
0.065&
0.157&
0.290\\
\hline
\end{longtable}
\end{table}

\begin{table}
Table 2. The production cross sections in fb for golden events at
LHC. The asterix {*} denotes that the calculations are performed assuming
\( m_{N}=50 \) GeV. \label{table2} 
\begin{longtable}{cccccc}
\hline 
\( m_{h} \)(GeV)&
SM-3&
SM-4&
SM-4{*}&
SM-5{*}&
SM-6{*}\\
\hline 
100&
0.19&
1.09&
1.09&
2.26&
2.89\\
110&
0.64&
3.48&
0.92&
2.32&
3.83\\
120&
1.93&
10.13&
1.52&
3.96&
6.95\\
130&
4.55&
24.74&
3.04&
8.04&
15.37\\
140&
6.57&
38.60&
4.95&
13.17&
23.91\\
150&
7.16&
47.77&
8.17&
21.77&
39.81\\
160&
3.31&
26.10&
10.61&
28.39&
52.61\\
170&
1.61&
13.10&
9.45&
25.60&
48.94\\
180&
3.58&
29.02&
22.85&
61.88&
118.16\\
190&
12.80&
102.82&
87.31&
235.53&
453.28\\
200&
14.40&
114.45&
99.14&
269.00&
517.89\\
220&
13.30&
102.70&
92.51&
249.69&
479.23\\
240&
12.40&
93.60&
85.99&
230.41&
440.72\\
260&
11.10&
80.85&
75.70&
201.26&
383.32\\
300&
9.60&
64.09&
61.08&
158.87&
298.71\\
350&
9.88&
53.42&
51.68&
127.54&
234.81\\
400&
8.30&
36.30&
35.55&
83.25&
149.36\\
450&
6.04&
25.26&
24.92&
58.59&
106.38\\
500&
4.23&
17.64&
17.43&
42.03&
77.52\\
\hline
\end{longtable}
\end{table}

\newpage

\end{document}